\documentclass[prb,twocolumn,superscriptaddress,shortbibliography]{revtex4-2}
\usepackage{graphicx}
\usepackage{morefloats}
\usepackage{color}
\usepackage{amssymb}
\usepackage{amsmath} 
\usepackage{float}
\usepackage{esvect}
\usepackage{wasysym}
\usepackage[normalem]{ulem}

\pdfoutput=1
\usepackage{pdfpages} %PDF PAGES
\makeatletter
\AtBeginDocument{\let\LS@rot\@undefined}
\makeatother

\usepackage[pagewise]{lineno}
\begin{document}

\title{Self-stabilizing exchange-mediated spin transport}

\author{T.\,Schneider}
%\email[]{t.schneider@hzdr.de}
\affiliation{Institute of Ion Beam Physics and Materials Research, Helmholtz-Zentrum Dresden - Rossendorf, 01328 Dresden, Germany}
\affiliation{Department of Physics, TU Chemnitz, Reichenhainer Str. 70, 09126 Chemnitz, Germany}
\affiliation{Department of Physics and Astronomy, University of California, Riverside, CA 92521, USA}
\author{D.\,Hill}
\affiliation{Department of Physics and Astronomy, University of California, Los Angeles, CA 90095, USA}
\author{A.\,K\'{a}kay}
\affiliation{Institute of Ion Beam Physics and Materials Research, Helmholtz-Zentrum Dresden - Rossendorf, 01328 Dresden, Germany}
\author{K.\,Lenz}
\affiliation{Institute of Ion Beam Physics and Materials Research, Helmholtz-Zentrum Dresden - Rossendorf, 01328 Dresden, Germany}
\author{J.\,Lindner}
\affiliation{Institute of Ion Beam Physics and Materials Research, Helmholtz-Zentrum Dresden - Rossendorf, 01328 Dresden, Germany}
\author{J.\,Fassbender}
\affiliation{Institute of Ion Beam Physics and Materials Research, Helmholtz-Zentrum Dresden - Rossendorf, 01328 Dresden, Germany}
\affiliation{Institute for Physics of Solids, Technische Universit\"at Dresden, Zellescher Weg 16, 01069 Dresden, Germany}
\author{P.\,Upadhyaya}
\affiliation{School of Electrical and Computer Engineering, Purdue University, West Lafayette, IN 47907 USA}
\author{Yuxiang\,Liu}
\affiliation{Department of Electrical and Computer Engineering, University of California, Los Angeles, CA 90095, USA}
\author{Kang\,Wang}
\affiliation{Department of Electrical and Computer Engineering, University of California, Los Angeles, CA 90095, USA}
\author{Y.\,Tserkovnyak}
\affiliation{Department of Physics and Astronomy, University of California, Los Angeles, CA 90095, USA}
\author{I.\,N.\,Krivorotov}
\affiliation{Department of Physics and Astronomy, University of California, Irvine, CA 92697, USA}
\author{I.\,Barsukov}
\email[]{igorb@ucr.edu}
\affiliation{Department of Physics and Astronomy, University of California, Riverside, CA 92521, USA}

%\keywords{superfluid, spin transport, Landau criterion, magnon instability, dipole interaction}
%\date{\today}
\begin{abstract}
Long-range spin transport in magnetic systems can be achieved by means of exchange-mediated spin textures with robust topological winding -- a phenomenon referred to as spin superfluidity. Its experimental signatures have been discussed in antiferromagnets which are nearly free of dipolar interaction. However, in ferromagnets, which possess non-negligible dipole fields, realization of such spin transport has remained a challenge.  Using micromagnetic simulations, we investigate coherent exchange-mediated spin transport in extended thin ferromagnetic films. We uncover a two-fluid state, in which the long-range spin transport by spin textures co-exists with spin waves, as well as a soliton-screened spin transport regime at high spin injection biases. Both states are associated with distinct spin texture reconstructions near the spin injection region and sustain spin transport over large distances.\\

This manuscript has been published in Phys. Rev. B 103, 144412 (2021)\\
https://link.aps.org/doi/10.1103/PhysRevB.103.144412
\end{abstract}
\maketitle

\section{Introduction}

The field of magnon-spintronics opens new possibilities for energy-efficient information storage\cite{FullertonSky,GOBEL2020}, transport, and processing. Achieving low-dissipation long-range spin transport is one of the main goals of spintronics research. In magnetic insulators, spin currents are carried by spin waves, free of undesired electric currents \cite{Chumak15}. However, despite low damping, spin waves exhibit exponential decay over distances that can be short at high frequencies.

The bosonic nature of spin excitations in ordered magnetic materials can benefit from magnon-magnon interactions \cite{Chris,3m} and the ensuing coherence. Bose-Einstein condensation of magnons, that was experimentally observed in various systems \cite{Oosawa1999,Nikuni2000,Radu2005,Demokritov2006,Bozhko2016}, is a notable example. Another phenomenon characteristic of bosonic systems is superfluidity; resistance-free charge transport in superconductors and viscosity-free mass transport in superfluid helium are some prominent examples \cite{Bardeen1957,Borovik-Romanov1984,Bennemann2014}. 

In Ref.\,\cite{Halperin1969}, Halperin and Hohenberg proposed a hydrodynamic theory of magnons, which is formally related to superfluidity. Extending this analogy further \cite{Sonin2010}, exchange-mediated spin transport by spin textures with metastable winding (EMS) can be dynamically induced in easy-plane ordered spin systems. Upon non-equilibrium spin injection with perpendicular-to-plane polarization \cite{Takei2014}, a global texture winding of magnetic order parameter develops in the form of a topologically robust winding spiral (Fig.\,1a). The order parameter precesses coherently in time at low frequencies, transporting spin over macroscopic distances \cite{Sonin2010} with slow algebraic decay \cite{Takei2014,Takei2015} governed by Gilbert damping \cite{Scheck, Bars, Nembach, YIGalpha}. The resultant spin transport is thus intrinsically long-ranged, beyond the decay length of ordinary spin waves. While EMS bears similarities to mass superfluidity (equation of spin motion resembles Josephson relations for superfluidity, superflow is characterized by the gradient of the phase)\cite{Koenig2001,Sonin2010,Takei2014,Chen2014,Flebus2016,Iacocca2017a,Iacocca2017,Tserkovnyak2017,Hill2018a,Hua2016,Sonin2018arxiv,Ezio2019,EzioReview,gonccalves2018oscillatory,Yizhou}, it must be stressed that this phenomenon is not truly dissipationless.

Recently, signatures of EMS have been experimentally observed in antiferromagnetic spin systems \cite{Stepanov2018,Yuan2018}. A realization of EMS in ferromagnets remains an unsolved challenge. Previous theoretical works have revealed the potential of EMS \cite{Koenig2001,Sonin2010,Takei2014,Chen2014,Flebus2016,Iacocca2017a,Iacocca2017,Tserkovnyak2017,Hill2018a,Hua2016,Sonin2018arxiv,Ezio2019,EzioReview,Yizhou} for spintronics applications but have not systematically studied the role of dipolar interactions. Numerical calculations in Ref.\,\cite{Skarsvag2015} for micrometer-scale thin-film ferromagnets have demonstrated that dipolar interaction can destroy EMS; moreover, numerical simulations in thin ferromagnetic stripes \cite{Iacocca2017,Iacocca2017a} have shown that EMS can be achieved despite the dipolar interaction, sparking a discussion on experimental feasibility of such states.

\begin{figure*}[t]
\centering
\includegraphics[width=0.99\textwidth]{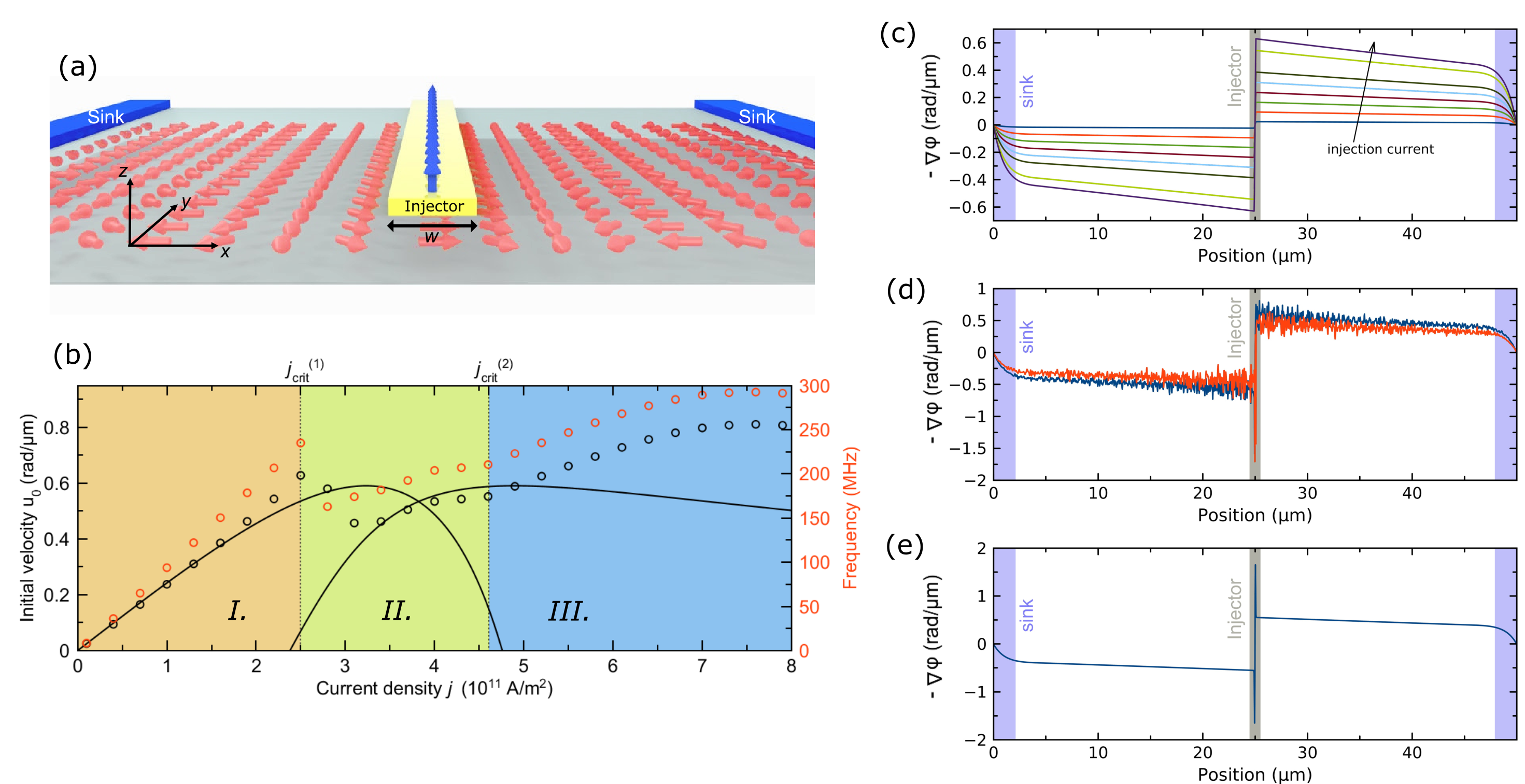}
\caption{EMS in the absence of dipolar interaction. (a)~Schematic view of the thin film. The spin injector provides spin current with out-of-plane polarization (blue arrows). The spin sinks are shown. The red arrows represent a magnetization snapshot. (b)~Initial EMS velocity (black circles) and base frequency (red circles) as functions of the current density. Three regimes of the EMS are marked. Black line shows transmitted spin current $\tau$ (in the same units as EMS velocity) calculated based on the analytical model.  (c)~Spatial dependence of the EMS velocity in regime I, (d)~in regime II at $j=3.1\cdot10^{11}$\,A\,m$^{-2}$,  (e)~in regime III at $j=4.6\cdot10^{11}$\,A\,m$^{-2}$.}
\label{fig1}
\end{figure*} 

Here we present a micromagnetic study of EMS in extended ferromagnetic thin films and investigate the role of dipolar interaction. We find that stable EMS state can in fact be achieved and that it allows for long-range spin transport in a wide range of spin biases.

In ferromagnetic films, the breakdown of the EMS corresponds to a sudden alignment of the magnetization out-of-plane, which disrupts the spin transport \cite{Sonin2010}. When the spin-injection bias is increased, the driven easy-plane spin-winding state reaches an instability at the injection site. We find that this, in turn, triggers a nonlinear dynamic state with significant out-of-plane character. By decreasing the injection efficiency, a stable easy-plane spin-winding state with a reduced spin flux can be recovered at some distance from the injector. We find that the spin bias applied to the injector does not determine the spin current flowing through the magnet. Instead, the latter is established self-consistently, as a result of the nonlinear feedback of the magnetic dynamics near the injector. This feedback regulates the spin injection through spin wave emission and/or coherent soliton formation. In effect, the spin superflow away from the injector can stay below the instability threshold even at large spin biases.

\section{Results}
We simulate extended ferromagnetic films in the thickness range of $d=2\text{--}30$\,nm by applying periodic boundary conditions in the film plane to a $50\,\mathrm{\mu m}\times5\,\mathrm{\mu m}$ patch. Magnetic parameters of the film are chosen (Appendix) to mimic Y\textsubscript{3}Fe\textsubscript{5}O\textsubscript{12} (YIG) -- a magnetic insulator with low damping that may be considered as a candidate for experimental realization of EMS. Magnetization dynamics is excited by locally injecting a continuous pure spin current with out-of-plane spin polarization. It is simulated through spin-transfer torque in the middle of the film underneath a narrow spin injector. The spin injector carries electric current \cite{PhysRevB.101.214412,PhysRevB.99.134409} that translates into spin current with conversion efficiency of $\theta_\mathrm{s}=0.07$ (Appendix). At the short edges of the film patch, spin sinks are simulated by a local increase of the Gilbert damping as explained in Appendix. The spin sinks are representative of spintronics devices fed and operated with spin current supplied through EMS. All calculations in this study are carried out at 0\,K, i.e. without thermal excitations. Figure~\ref{fig1}a shows the sample geometry, spin injector, and spin sinks.\\

\textbf{Behavior without dipolar interaction.} At first, we investigate the case of omitted dipolar interaction by enforcing zero dipole fields in our simulations and introducing an artificial easy-plane anisotropy $K_{\mathrm{u}}=-10$\,kJ\,m\textsuperscript{-3} approximating the shape anisotropy of a thin film \cite{Kravchuk2018, Takei2014, barsukovfielddep}. For each current value, the simulations are carried out until steady state or dynamic equilibrium is reached. In Fig.~\ref{fig1}a, a snapshot of magnetization is shown for the steady state at a current density $j=10^{11}$\,A\,m\textsuperscript{-2} in the spin injector. The magnetization exhibits continuous  2$\pi$-rotations in the film plane, characteristic of the EMS state \cite{Sonin2010}. The EMS velocity is defined as $u(x)=-\nabla \phi(x)$, where $\phi$ is the azimuthal angle of magnetization \cite{Sonin2010, Takei2014} (it is also the local order parameter of the EMS state). Figure~\ref{fig1}b shows the initial velocities $u_0$ (calculated in the vicinity of the injector region as described in Appendix) as a function of the current density. Three distinct regimes can be identified as indicated in the figure:

\textit{Regime I}. At low current densities, the EMS velocity increases linearly with increasing current density, in good agreement with analytical predictions of Ref.~\cite{Upadhyaya2017}. The EMS velocity decreases smoothly and slowly with increasing distance from the spin injector (Fig.\,\ref{fig1}c). At the spin sink, it decreases more rapidly and reaches zero value. The longitudinal spin density $n=m_z$ (equal to the polar component of the normalized magnetization) \cite{Iacocca2017a} is well below 0.5.

{\textit{Regime II}.} At the first critical current density $j_\mathrm{crit}^{(1)}$, the EMS starts to exhibit oscillations in real space, as shown in  Fig.\,\ref{fig1}d. The initial EMS velocity is calculated by averaging out these oscillations. It shows a notable drop at the first critical current (Fig.\,1b). Underneath the injector, the magnetization is partially tilted out of the film plane by the spin current. Outside of the injector region, the longitudinal spin density remains $n<0.5$.%; the system stays in the subsonic regime.

Analysis of the temporal evolution of magnetization reveals large oscillations in the injector region. It emits incoherent spin waves into the rest of the film which superimpose with the EMS state (Fig.\,1d). We observe spin wave emission and the drop of the EMS velocity for various injection widths $w=30\text{---}300$\,nm. The injector width does not affect the critical current, but modifies \cite{Iacocca2017} the critical current density through geometrical renormalization $j^{(1)}_\mathrm{crit}\propto I_\mathrm{crit} / w$ (Fig.\,4 in Appendix).

The temporal base frequency $\Omega$ of the EMS spiral is extracted for each current density by calculating the fast-Fourier transformation of the time evolution of the in-plane magnetization components in the injector region. The frequency governs the Gilbert dissipation of the EMS as $\alpha \Omega^2$, where $\alpha$ is the Gilbert damping constant (Appendix). Moreover, it is a temporal characteristic of the EMS spin dynamics that can be observed in experiment. As shown in Fig.\,\ref{fig1}b, both $u_0$ and $\Omega$ exhibit the distinct breakdown in the regime II.

{\textit{Regime III}.} Above the second critical current density $j_\mathrm{crit}^{(2)}$, the EMS velocity is again a smooth function of the distance (Fig.\,1e). No spin waves are observed. The magnetization underneath the injector is almost fully aligned out-of-plane and does not vary with time. Both initial velocity and base frequency show a reduced growth rate with increasing spin current and saturate around $ j = 8 \cdot 10^{11}$\,A\,m\textsuperscript{-2} (Fig.\,1b).\\

\textbf{Analytical model.} We strive to develop \cite{PRE} a minimal analytical model to explain the observed phase diagram; we thus neglect dipolar interaction and magnetic damping. With exchange constant $A_{\mathrm{ex}}$, we employ the free energy:
\begin{equation}
F= \int \mathrm{d}x^3 \left[ A_{\mathrm{ex}} \left(\nabla \pmb{m}\right)^2 - K_\mathrm{u} m_z^2 \right].
\end{equation}
Taking into account that magnetization $\pmb{m}$ does not vary along the $y$ and $z$ directions, Landau-Lifshitz equation assumes the form
\begin{equation}
 \frac{\mathrm{d}\pmb{m}}{\mathrm{d}t} = -  \pmb{m} \times \left(\frac{\partial^2 \pmb{m}}{\partial x^2} - m_z \pmb{\hat{z}}\right),
 \label{eqn:LLG2}
\end{equation}
where $x$ and $t$ are re-scaled in units of $\sqrt{A_{\mathrm{ex}}/K_\mathrm{u}}$ and $\mu_0 M_{\mathrm{s}}/{2\gamma K_\mathrm{u} }$, respectively (with the permeability of free space $\mu_0$ and gyromagnetic ratio $\gamma$). By parameterizing the magnetization with spherical coordinates, $\pmb{m}=( \text{sin}\, \theta \, \text{cos}\,\phi, \text{sin}\, \theta \,\text{sin}\, \phi, \text{cos}\,\theta)$, equation (\ref{eqn:LLG2}) becomes
\begin{equation}
\dot{\theta} \, \text{sin}\,\theta = -\partial_x (\text{sin}^2 \theta \,\partial_x \phi),
\end{equation}
\begin{equation}
\dot{\phi} \, \text{sin} \, \theta  = \partial_x^2 \theta +\frac{1-(\partial_x \phi )^2}{2} \text{sin} \, 2\theta.
\end{equation}
Equation (3) corresponds to a continuity equation for the longitudinal spin density. We are interested in solutions which satisfy boundary conditions of the form
\begin{equation}
-\partial_x \phi(0) = \tau_i - \tilde{\gamma} \partial_t \phi(0), \;\;\; -\partial_x \phi(L) = \tilde{\gamma} \partial_t \phi(L)
\label{eqn:BCS}
\end{equation}
where $\tau_i$ is the spin torque from the injection site and $\tilde{\gamma}$ parameterizes the edge damping effects of spin pumping \cite{Hill2018a, Brataas2002}. General soliton solutions of equation (\ref{eqn:LLG2}) were studied in Ref.\,\cite{PRE}. Here we develop soliton solutions with boundary conditions (\ref{eqn:BCS}).
Assuming soliton solutions have the form $\theta=\theta(x-c t)$ results (Appendix) in
\begin{equation}
\phi-\phi_0 = \Omega t - \int_{0}^x \text{d}x' \frac{c \, \text{cos}\, \theta +a_1}{\text{sin}^2\theta},
\end{equation}

\begin{equation}
x-ct=x_0\pm\frac{1}{\sqrt{2}}\int_{\theta_1}^{\theta(x,t)} \frac{\text{d}\theta'}{\sqrt{f(\theta')}},
\end{equation}
where $f(\theta)=a_2- \Omega \cos \theta - \frac{1}{2} \text{sin}^2  \theta - \frac{1}{2} (c^2 - a_1^2)\csc ^2(\theta) - ca_1 \cot (\theta) \csc (\theta) $. Here $\Omega$, $c$, $\phi_0$, $a_1$, and $a_2$ are integration constants. We consider the case in which $f(\theta)>0$ for some open interval $(\theta_1,\theta_2) \subset (0,\pi/2) $, where $\theta_1$ and $\theta_2$ are zeros of $f(\theta)$. The soliton expression (7) results in a solution of the form $(x-ct)(\theta)$, which is multi-valued, i.e. it has multiple branches that need to be pieced together to obtain the inverted result of the form $\theta(x-ct)$. In this patching procedure, solution branches are selected that produce physically meaningful results. The procedure is guided by the results of the micromagnetic simulations and is carried out in compliance with the spatial continuity of magnetization and its first derivative, as well as satisfying the boundary conditions (injector and sink). The resulting soliton solution, $\theta(x-ct)$, is symmetric about its minimum $\theta_1$ (corresponding to a spike in $m_z$) centered at $x_0$ at $t=0$. One such solution is an isolated soliton traveling at speed $c$ through a surrounding EMS which has constant polar angle $\theta_2$. The length of the soliton is determined by the characteristic length scale $\sqrt{A_{\mathrm{ex}}/K_\mathrm{u}}$. 

%only describes $\theta$ over a finite interval of $x-ct$, to describe a full solution, the soliton and its first derivatives must be patched to suitable surrounding solutions, such as another soliton or a EMS, or to the boundary conditions. This patching process fixes the integration constants. 

The analytically calculated transmitted spin current per spin density ($\tau = -\nabla \phi \,\sin^2\theta$) is shown in Fig.\,1b as the black solid line. The analytical spin current plot shows three distinct phases, similar to the three phases identified in micromagnetic simulations. The low-current regime (I) corresponds to the conventional EMS, i.e. a coherently precessing constant-$\theta$ superflow as derived in Ref.\,\cite{Tserkovnyak2017}. 

For high injection currents of the regime (III), on the other hand, we find a stationary soliton solution $c=0$ of particular interest. The soliton is placed with the peak at the injection region boundary. The injector region is nearly fully polarized out-of-plane $\theta \to 0$, and the local time-dependent oscillations in $\theta$ cease. As micromagnetic simulations show, this configuration lacks the spin wave noise. The spin current is reduced by the injector edge soliton due to the high out-of-plane magnetic polarization near the injector. This configuration with high out-of-plane polarization diminishes the transmitted spin current at the same EMS velocity $u$. The polarization in the injector region partially blocks the spin injection, and the transmitted spin current asymptotically behaves as $\propto 1/j$ for $j \rightarrow \infty$. By virtue of this self-regulation in the injector region, the EMS persists above biases expected for the instability (for this reason, we refer to it as screened EMS or screened spin superfluid).

In a previous analytical study \cite{Tserkovnyak2017}, the drop of the transmitted spin current to zero after the first critical current has been associated with EMS becoming fully polarized out of plane ($\theta=0$). However, such state is in fact unstable, even in the undamped model. It has a mode of instability which forms near the boundaries and propagates into the rest of the film. This mode of instability has EMS-like precession and grows exponentially with time. Furthermore, the micromagnetic simulations suggest that neither the conventional EMS of regime (I) nor the screened EMS of regime (III) are stable when the two independent solutions of the analytical model overlap in regime (II) (Fig.\,1b). Instead, the simulations show that EMS persists in the form of a non-trivial dynamic state. The solution may be a hybrid periodically transitioning between the conventional EMS and screened EMS. This transitioning results in injector region oscillations and spin waves propagating into the film.

\begin{figure}[b]
\centering
\includegraphics[width=0.99\columnwidth]{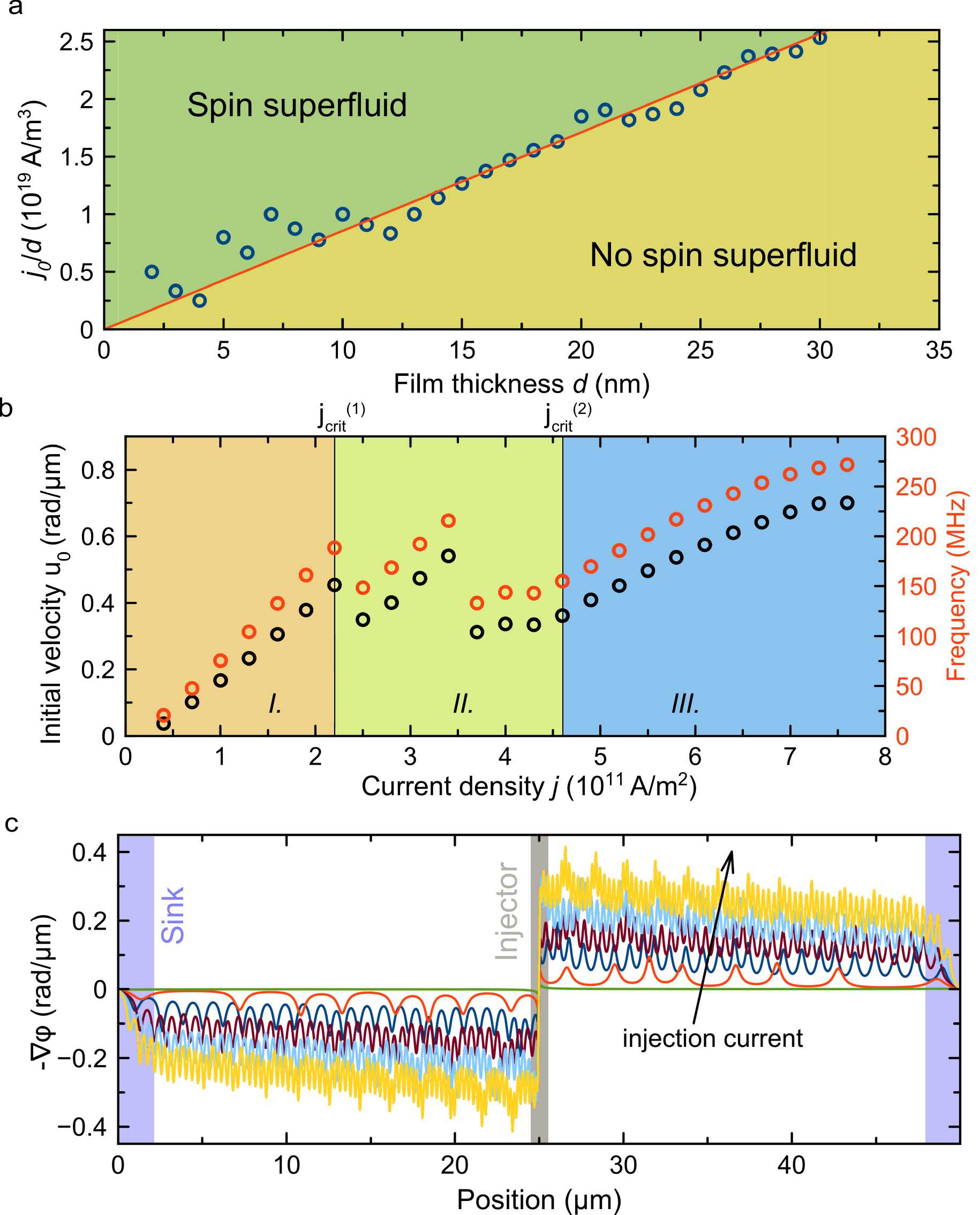}
\caption{EMS in the presence of dipolar interaction. (a)~Threshold current as a function of film thickness. (b)~Initial EMS velocity and base frequency as functions of the current density for a 5\,nm thick film (sub-threshold regime omitted for clarity). (c)~Spatial dependence of the EMS velocity.}
 \label{fig2}
\end{figure}

\textbf{Impact of dipolar interaction.} A previous study \cite{Skarsvag2015} on micron-sized ferromagnetic thin films has pointed out a detrimental effect of the dipolar interaction on the EMS, leading to a collapse of the operable bias range. Here, we investigate extended systems by employing periodic boundary conditions. In the following micromagnetic simulations, the dipolar interaction is enabled and the previously used uniaxial anisotropy $K_\mathrm{u}$ is set to zero.

First, we find that the presence of the dipolar interaction suppresses EMS at low currents and imposes a threshold $j_0$ for its formation \cite{Skarsvag2015}. The uniaxial anisotropy $K_\mathrm{u}$, introduced in the previous simulations to mimic the shape anisotropy, has enabled a simple easy-plane magnetic system in which the EMS can form without injection threshold. On the other hand, the nonlocal nature of the dipolar interaction introduces an effective magnetic anisotropy -- an energy barrier to overcome -- for the formation of the spatially periodic spin texture of the EMS  \cite{Sonin2010,Iacocca2017a,Iacocca2017}. The effective dipole energy increases with the thickness of the film $d$, which is varied between 2\text{--}30\,nm in our simulations. For comparison across different film thicknesses, the current needs to be scaled by $d$. Indeed, Fig.\,2a shows that a normalized threshold current $j_0/d$ increases nearly linearly with increasing film thickness. 

Upon the formation of the EMS, its initial velocity $u_0$ shows non-monotonous dependence on the current density. Figure~2b shows a behavior qualitatively similar to omitted dipolar interaction. Employing spatio-temporal analysis of the magnetization dynamics, we find again: (I) the low-current regime free of incoherent spin waves, (II) the intermediate regime with co-existing EMS and incoherent spin waves, and (III) the high-current regime of screened EMS, free of incoherent spin waves. An additional notable drop of the initial velocity and base frequency is observed in the middle of the intermediate regime (II). A detailed evaluation of the data reveals that $u_0$ and $\Omega$ show multiple non-monotonicities for both the dipole and dipole-free cases. While the currents at which they occur differ, their presence seems to be universal and is likely related to the non-linear generation of spin waves in regime II.

We further find differences of the spatial profile of EMS velocity compared to the dipole-free case. As shown in Fig.\,2c, the gradient of the azimuthal angle exhibits spatial modulations. Due to the continuous rotations of magnetization, dipolar interaction introduces perturbations of the energy landscape with uniaxial symmetry -- the dipolar field alternates at every $\pi$-rotation. The in-plane components of the magnetization (Fig.\,3a) display a distorted sinusoidal profile as a function of distance. Thus, the angle gradient shows a small magnitude modulation with the period of the $\pi$-rotations. The out-of-plane component of magnetization reveals spikes at the extrema of $m_x$ (Fig.\,3), which reduces the exchange energy. This modulation can be considered a soliton lattice, resulting in an EMS state with a broken symmetry. The symmetry is broken by the spin injector and mediated to the EMS by virtue of the dipolar interaction.

\begin{figure}
\includegraphics[width=\columnwidth]{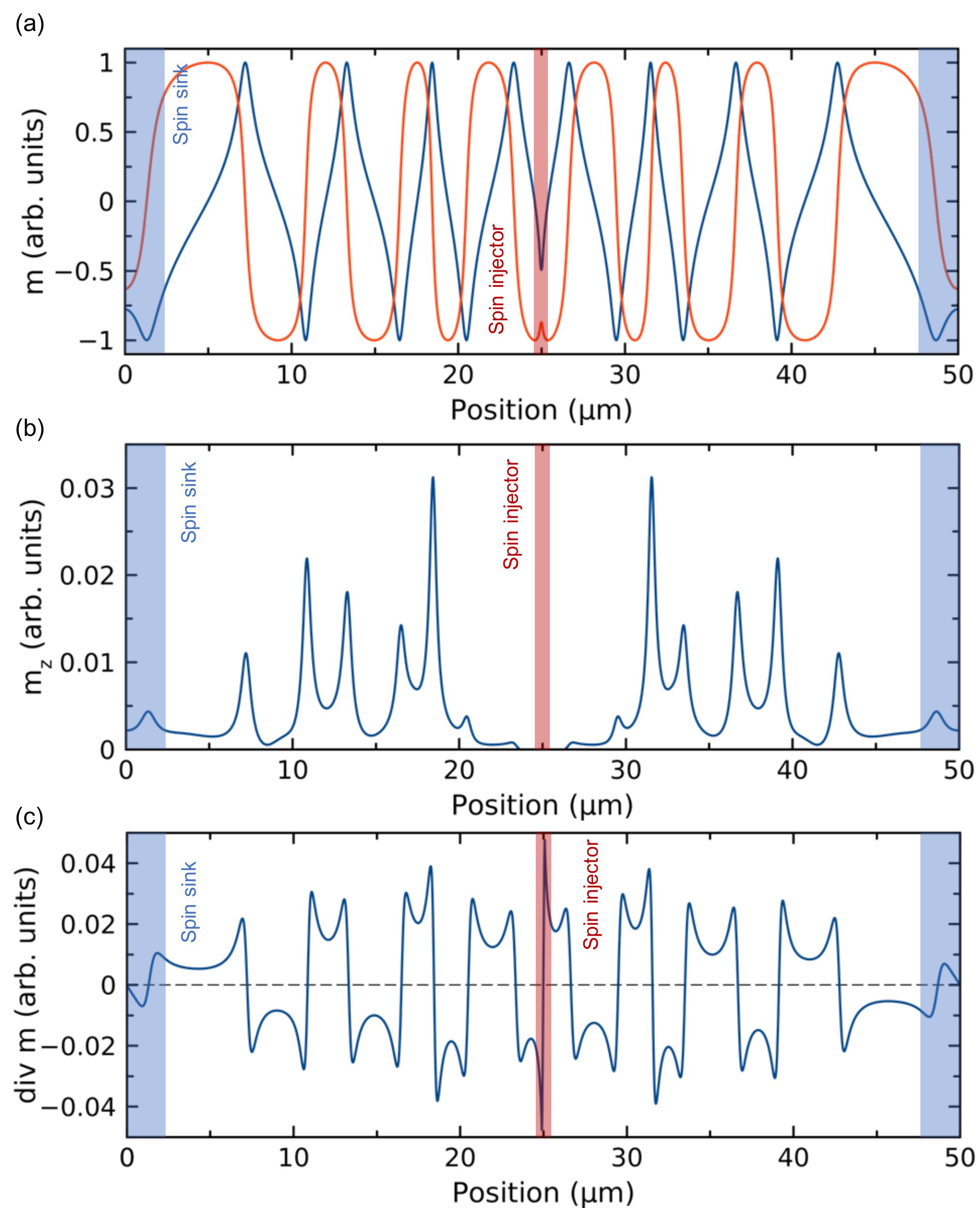}
\caption{Perturbations of the magnetization spiral in the presence of dipolar interaction, snapshot after 500\,ns for $j=4\cdot10^{10}$\,A\,m$^{-2}$. (a)~In-plane components of the normalized magnetization, $m_{\mathrm{x}}$ (blue solid line) and $m_{\mathrm{y}}$ (red solid line), deviate from the sinusoidal behavior. (b)~Out-of-plane component $m_{\mathrm{z}}$ reveals peaks when $m_{\mathrm{x}}$ has an extremum. (c)~The divergence of the magnetization is linked to the magnetostatic field.}
\label{SuppFig2}
\end{figure}

\section{Conclusions}
In this study, the EMS is found to persist over a large range of bias currents. The magnetization pinning \cite{BarsukovOxide} by the dipole fields  \cite{Skarsvag2015} does not fully suppress the EMS at high biases for the case of extended films \cite{Iacocca2017a}. The threshold suppression of the EMS at low biases has been previously discussed \cite{Sonin2010,Hill2018a} for symmetry-breaking magnetic anisotropy. In contrast to the effect of such local anisotropy, the symmetry breaking, investigated in this study, is mediated by the non-local dipolar interaction \cite{Iacocca2017}. We find the threshold current to increase linearly with increasing dipole energy.

A coupling between the EMS order parameter (azimuthal angle $\phi$) and the longitudinal spin density $n$ is observed. The longitudinal spin density shows oscillations at twice the base frequency \cite{Iacocca2017}, in agreement with the symmetry order of the effective (uniaxial) magnetic anisotropy due to the dipole fields. The oscillations correspond to excitations of the soliton lattice. No such behavior is observed in the absence of the dipolar interaction.

We identify three regimes of the EMS, universally present, with and without dipolar interaction. In the low-current regime, conventional EMS is found. Above the first critical current, the EMS co-exists with incoherent non-thermally populated magnons. Above the second critical current, the incoherent magnons are suppressed and a soliton-screened EMS is found.

We discover that the EMS can self-stabilize beyond the anticipated critical injection bias. The spin superflow is not determined by the injection current alone but self-consistently, taking into account the spin reconstruction in the injector region. At high biases, the EMS is partially screened from injected spin current by soliton formation. For the intermediate-current regime, we identify non-linear magnon scattering to play a role in EMS self-stabilization.

Recently, spin injection with perpendicular polarization due to spin-orbit effect with spin rotational-symmetry \cite{Xiao} and due to planar Hall effect \cite{Ilya} has been experimentally realized using metallic ferromagnets. Moreover, efficient thermal spin injection \cite{Tana} with polarization not bound by injector geometry has been achieved \cite{Chris,arkook2019thermally}. These developments may benefit designing novel ferromagnetic spin injectors and instigate research on thin film-based EMSs. Questions on spin texture formation in the injector region due to interaction with the injector, thermal stability of the superflow, and accessible spin bias ranges are likely to arise. Our work points out the impact of injector spin texture formation and incoherent spin waves on stabilization of EMS and extending the range of achievable spin biases.

\section*{APPENDIX}
\textbf{Micromagnetic simulations.} Micromagnetic simulations were carried out by numerically solving the Landau-Lifshitz equation using MuMax software \cite{Vansteenkiste2014}. The sample volume was discretized into a mesh with the cell size of 24.41\,nm$\times$19.53\,nm$\times d$. The validity of the results was validated by carrying out control simulations with a reduced cell size. Periodic boundary conditions within the MuMax code were used. For the dipole-free simulations, dipolar interaction was disabled within the code. For the simulations with the dipolar interaction, the magnetostatic field was accounted for using the approach presented in Ref.\,\cite{Wang2010}. 

The spin current injection was simulated via the spin-transfer (Slonczewski) torque within the code. The electric current density given throughout the manuscript corresponds to the injected spin current via $j_{\mathrm{s}} = \theta_{\mathrm{s}} \frac{\hbar}{e} j$ with the spin conversion efficiency $\theta_{\mathrm{s}}$, the Planck constant $\hbar$ and the elementary charge $e$.

The spin sinks were designed to emulate spintronic devices, fed by the spin current transmitted through the EMS. They were modeled by non-uniform increase of the Gilbert damping over the width (4\,$\mu$m) of the spin sink regions. From the sink edge closer to the injector to the edge at the end of the film patch, the damping constant $\alpha$ was increased exponentially from 0.002 to 0.11. Such modification of damping emulates extraction of angular momentum from the EMS and prevents potential reflections of spin excitations at the edges of the patch, which is necessary to simulate an extended spin system.

In the simulations, the magnetization dynamics was evolved for several hundreds of nanoseconds. The data shown in the figures corresponds to snapshots of the spatially dependent magnetization. The snapshots were taken after the magnetization field had reached the dynamic steady-state. Typically, a simulation time of $500$\,ns was chosen to ensure that the system reached the steady-state. Control calculations with various simulation times were performed to ensure the validity of the steady-state.

The position dependent EMS velocity $u(x)=-\nabla \phi(x)$ was calculated from the position dependent azimuthal angle $\phi$, determined from the magnetization snapshots after reaching the steady-state. The initial EMS velocity $u_0$ was obtained from the position dependent EMS velocity in the direct vicinity to the injector edge (and by averaging out the spatial modulations in this region). The base frequency was determined as a maximum-amplitude frequency of the Fourier transformation of the time dependent in-plane magnetization data within the injector region.

The material parameters were chosen to simulate YIG films \cite{Solt1962,Klingler2014,Dubs2017,MewesPMA}: the saturation magnetization $M_{\mathrm{s}} = 130\,\mathrm{kA\,m^{-1}}$ and the exchange constant $A_{\mathrm{ex}}  = 3.5\,\mathrm{pJ\,m^{-1}}$. Control simulations with the magnetocrystalline anisotropy of YIG (0.6\,kJ/m$^3$) were carried out. The anisotropy results in an increase of the threshold current consistent with previous studies \cite{Iacocca2017}. This effect is small compared with the threshold induced by the dipolar interaction. Moreover, magnetocrystalline anisotropy induces a spatial modulation of the EMS velocity (with smaller wavelength than the dipolar modulation of the velocity) that is consistent with the magnetocrystalline anisotropy symmetry \cite{Iacocca2017}. This modulation, together with the modulation due to the dipolar interaction, leads to an overall complex spatial profile of the EMS velocity. As it does not contribute to the discussion of this study, the magnetocrystalline anisotropy was omitted in the simulations. 

\begin{figure}[b]
\includegraphics[width=\columnwidth]{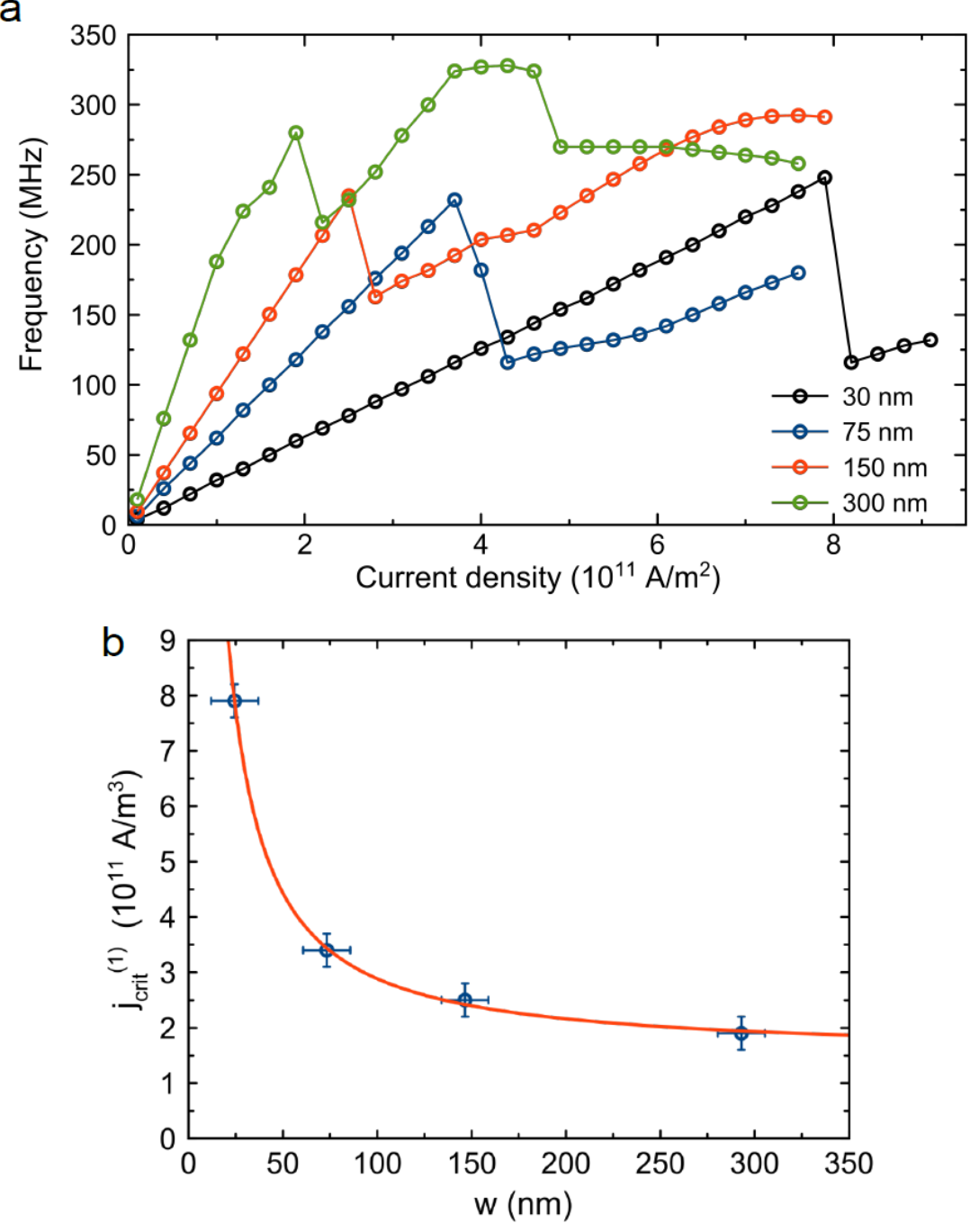}
\caption{Impact of the injector width on EMS. (a) Base frequency $\Omega$ for different injector widths. (b) The first critical current density decreases as $\propto 1/w$ (red line), where $w$ is the injector width.}
\label{SuppFig1}
\end{figure}

\textbf{Analytical model.} Numerical calculations of the analytical model resort to the same material parameters as micromagnetic simulations, but do not include magnetic damping. The parameters of the analytical model, spin sink (edge damping $\tilde{\gamma}$) and spin conversion efficiency at the interface, were fixed by fitting the low-bias analytical model to the results from the micromagnetic simulations. 

Here we derive equations (5) and (6). The assumption $\theta = \theta(x-ct)$ implies that the left-hand-side of Equation~(3) can be written as a derivative in $x$, thus allowing Equation~(3) to be integrated. The result can be solved for $\partial_x\phi$ and integrated again to express $\phi$ in terms of $\theta$. In general, the constants of integration can depend on $t$, i.e.
\begin{equation}
\phi =C_2(t) -\int^x dx' \frac{c \, \text{cos}\,\theta+C_1(t)}{\text{sin}^2\theta}. 
\end{equation}

However, the time dependence is restricted by substituting the expression for $\phi$ in terms of $\theta$ into Equation~(4). Once $\theta$ has been isolated, the resulting equation should not have explicit $t$ dependence because, by assumption, $\theta$ only depends on $x-ct$. This implies that $C_1$ is independent of time and restricts $C_2$ to at most linear dependence on $t$, thus resulting in equation (6). Once $\phi$ dependence has been eliminated in Eq.\,(4), equation (7) follows by direct integration with the integrating factor $\partial_x\theta$.

\section*{Acknowledgments} 
Investigation of the dipolar effects was supported as part of the "Spins and Heat in Nanoscale Electronic Systems" (SHINES), an Energy Frontier Research Center funded by the U.S. Department of Energy, Office of Science, Basic Energy Sciences (BES) under Award \#\,SC0012670. Investigation of the EMS phase transitions was supported by the National Science Foundation under Grant No.~ECCS-1810541. The analytical modeling by D.H. and Y.T. was supported by the US Department of Energy, Office of Basic Energy Sciences, under Award No.~{DE-SC0012190}. We thank NVIDIA Corporation for the donation of Titan Xp GPU used for some of the calculations.  We thank J\"urgen K\"{o}nig, Se Kwon Kim, and Hector Ochoa for helpful discussions.

%\bibliography{SpinSuperfluid.bib}
%\bibliographystyle{h-physrev}

\end{document}